\documentclass{iaus}
\usepackage{graphicx}
\usepackage{natbib}

\def\nodata{...}

\def\km/s{km~s$^{-1}$}

\def\Msun{\hbox{\it M$_\odot$}}

\def\Lsun{\hbox{\it L$_\odot$}}
\def\Myr{Myr}

\def\Minit{\hbox{M$_{\rm initial}$}}

\def\FMM362{FMM362}
\def\mnras{{\it Monthly Notices of the Royal Astronomical Society\, }}

\def\aap{{\it Astronomy \& Astrophysics\, }}
\def\nat{{\it Nature\, }}
\def\apj{{\it Astrophysical Journal\, }}

\def\aj{{\it Astronomical Journal\, }}
\def\pasp{{\it Publications of the Astronomical Society of the Pacific\, }}
\def\apjl{{\it Astrophysical Journal Letters\, }}

\def\simgr{\mathrel{\hbox{\rlap{\hbox{\lower4pt\hbox{$\sim$}}}\hbox{$>$}}}}
\def\simls{\mathrel{\hbox{\rlap{\hbox{\lower4pt\hbox{$\sim$}}}\hbox{$<$}}}}

\long\def\symbolfootnote[#1]#2{\begingroup%
\def\thefootnote{\fnsymbol{footnote}}\footnote[#1]{#2}\endgroup} 

\title[JD 11.~~Young Massive Clusters] 
{Young Massive Clusters}

\author[Donald F. Figer]   
{Donald F. Figer$^1$
}

\affiliation{$^1$Chester F. Carlson Center for Imaging Science\\
Rochester Institute of Technology\\
Rochester, NY 14623-5604 \\email: {\tt figer@cis.rit.edu}}

\pubyear{2008}
\volume{250}  
\pagerange{119--126}
\jname{Massive Stars as Cosmic Engines}
\editors{F. Bresolin, P.A. Crowther \& J. Puls, eds.}
\begin{document}

\maketitle

\begin{abstract}
Over the past ten years, there has been a revolution in our understanding of massive young stellar clusters in the Galaxy. 
Initially, there were no known examples having masses $>$10$^4$, yet we now know that there are at least a half dozen 
such clusters in the Galaxy. In all but one case, the masses have been determined through infrared 
observations. 
Several had been identified as clusters long ago, but their massive natures were only recently determined. 
Presumably, we are just scratching the surface, and we might look forward to having statistically significant 
samples of coeval massive stars at all important stages of stellar evolution in the near future. I review the efforts that have 
led to this dramatic turn of events and the growing sample of young massive clusters in the Galaxy.
\end{abstract}

\firstsection 
\section{Introduction}
Massive stellar clusters are the birthplaces of massive stars. They are also the places where many massive
stars reside all of their lives, having little time to wander before exploding as supernovae. The
astrophysical importance of massive clusters largely derives from their content of massive stars, objects
that have extraordinary effects on their surroundings. 
Indeed, massive stars are key ingredients and probes of astrophysical phenomena on all size and
distance scales, from individual star formation sites, such as Orion, to the early Universe during
the age of reionization when the first stars were born. As ingredients, they control the dynamical
and chemical evolution of their local environs and individual galaxies through their influence on
the energetics and composition of the interstellar medium. They likely play an important role in
the early evolution of the first galaxies, and there is evidence that they are the progenitors of the
most energetic explosions in the Universe, seen as GRBs. As probes, they define the upper limits
of the star formation process and their presence may end further formation of nearby lower mass
stars and planets. 

Despite the importance of massive stars, and the clusters in which they reside, no truly massive clusters 
were known to exist in the Galaxy before about ten years ago, when the Arches and Quintuplet clusters were identified
as being at least that massive \citep{fig99b}. Since then, a number of efforts have led to the identification
of about a half dozen more such massive clusters in the Galaxy. 

The evolution of massive stars is difficult to study because many of the most important phases
are short. To date, one of the most effective techniques to overcome this problem is to identify
massive clusters at ages when its most massive stars are largely in a single phase of evolution.,
We see this in the Arches cluster (the most massive stars, i.e.\
hydrogen burning Wolf-Rayet stars on the main sequence), 
The Central Cluster (Ofpe/WN9 stars), Westerlund~1 (Wolf Rayet stars), and the Scutum red supergiant clusters (RSGs).

We are on the cusp of a revolution in massive stellar cluster research, as the identified sample 
is likely the ``tip of the iceberg.'' In the next ten years, we can expect to identify perhaps a
factor of ten more massive clusters in the Galaxy than are currently known. With this sample, we can
expect to routinely address many of the long-pursued questions in massive star research to
determine, for example: 1) the most massive star that can form, 2) the binary frequency of massive
stars, 3) the properties of massive stellar winds, e.g.\ clumping, 4) the evolutionary sequence of
massive stars, and 5) the end states of massive stars. Indeed, \citet{bar04} summarize many interesting
questions that might be addressed with future studies of massive stars. 

In this paper, I review the Galactic sample of massive stellar clusters (M$>$10$^4$~\Msun), the efforts that led to their
discovery, and the identification of a stellar upper mass limit. Finally, I speculate on the hidden
population of massive stellar clusters in the Galaxy that will likely be revealed in the near future.

\section{The sample of massive young clusters}

There are approximately ten known Galactic clusters with masses $\gtrsim$10$^4$~\Msun, with three being located within
the central 50~pc of the Galactic center. Table \ref{proptable} gives the known properties of these clusters
in a list that is ordered with decreasing mass.
In some cases, the properties are poorly determined, in which case the entries are filled with an ellipsis.
The most massive, Westerlund~1, has a mass of $\approx$50,000~\Msun. The densest are in the Galactic center,
with the Arches cluster having a density of $\approx$300,000~\Msun~pc$^{-3}$. While most are quite young,
the RSG (red supergiant) clusters are distinctly older. Presumably, the table reflects an observational bias in that
younger clusters are generally more concentrated, and thus easier to identify as clusters. 

In the following, I review some of the clusters with more well-established characteristics.

\begin{table}
\scriptsize{
\begin{center}
\caption{Properties of massive clusters in the Galaxy$^a$\label{proptable}}
\begin{tabular}{lrrrrrrrrrrrrr}\hline
&
Log(M) &
Radius &
Log($\rho$) &
Age &
Log(L) &
Log(Q) &
OB &
YSG &
RSG &
LBV &
WN &
WC
\\ 
Cluster &
\Msun &
pc &
\Msun \, pc$^{-3}$ &
\Myr &
\Lsun & 
s$^{-1}$ &  
 &  &  &  &  & 
\\ \hline
Westerlund 1$^b$ & 4.7& 1.0& 4.1& 4$-$6& \nodata   & \nodata   & \nodata & 6 & 4& 2 & 16& 8 \\
RSGC2$^c$ & 4.6& 2.7& 2.7& 14$-$21& \nodata  & \nodata  & 0 & 0 & 26& 0& 0& 0\\
RSGC1$^d$ & 4.5& 1.3& 3.5& 10$-$14& \nodata  & \nodata  & 1& 1& 14& 0& 0& 0\\
Quintuplet$^e$ & 4.3& 1.0& 3.2& 4$-$6& 7.5 & 50.9 & 100 & 0 & 1 & 2 & 6 & 13 \\
Arches$^f$& 4.3& 0.19 & 5.6& 2$-$2.5& 8.0& 51.0 & 160 & 0 & 0 & 0 & 6 & 0 \\
Center$^g$& 4.3& 0.23& 5.6& 4$-$7& 7.3& 50.5 & 100 & 0 & 4 & 1 & 18 & 12 \\
NGC 3603$^h$ & 4.1& 0.3& 5.0& 2$-$2.5& \nodata  & \nodata  & 60 & 0 & 0& 0& 3& 0\\
Trumpler 14$^i$ & 4.0& 0.5 & 4.3 & $<$2 & \nodata & \nodata& 31& \nodata& \nodata& \nodata&  \nodata&  \nodata \\
Westerlund 2$^j$ & 4.0& 0.8 & 3.7& 1.5$-$2.5  & \nodata  & \nodata  & \nodata  & \nodata  & \nodata & \nodata  & 2  & \nodata  \\
Cl 1806-20$^k$ & 3.8& 0.8& 3.5& 4$-$6& \nodata  & \nodata  & 5 & 0 & \nodata & 1 & 2 & 2 \\ \hline
\end{tabular}
\end{center}
\vspace{1mm}
 \scriptsize{
   $^a$An ellipsis has been entered in cases where data are not reliable or available. 
``M'' is the total cluster mass in all stars extrapolated down to a lower-mass cutoff of 1 \Msun, 
assuming a Salpeter IMF slope and an upper mass cutoff of 120 \Msun\ (unless otherwise noted) ``Radius'' 
gives the average projected separation from the centroid position. ``$\rho$'' is M divided by the 
volume. This is probably closer to the central density than
the average density because the mass is for the whole cluster while the radius is the average projected 
radius. ``Age'' is the assumed age for the cluster. ``Luminosity'' gives the 
total measured luminosity for observed stars. ``Q'' is the estimated Lyman continuum flux emitted by the cluster. 
$^b$\citet{fig06}. 
$^c$\citet{dav07}. 
$^d$\citet{fig06}. 
$^e$\citet{fig99b}. 
$^f$Mass estimates have been made based upon the number of stars having \Minit$>$20~\Msun\ given in \cite{fig99b} and 
the mass function slope in \citet{kim06}. The age, luminosity and ionizing flux are from \citet{fig02}.
$^g$\citet{kra95}. The mass, M has been estimated by assuming that a total 10$^{3.5}$ stars have been formed. 
The age spans a range covering an initial starburst, followed by an exponential decay in the star formation rate.
$^h$\citet{har07}. 
$^i$\citet{har07}. 
$^j$\citet{har07}. 
$^k$\citet{fig05a}.}
}
\end{table}

\subsection{Westerlund~1}
Westerlund~1 is the most massive young cluster known in the Galaxy \citep{cla05,neg05,ski06,gro06,cro06,bra08}.
Given its age of $\approx$4~\Myr, it contains more evolved massive stars 
than any other cluster in the Galaxy, including half the known population of yellow
supergiants and the most WR stars. The cluster contains a magnetar, a highly magnetized
neutron star that may be descended from a particularly massive progenitor \citep{mun06}.
Oddly, the massive stellar content in Westerlund~1
has only been recently revealed, more than 40 years after the cluster's discovery. 

\subsection{Red Supergiant clusters}
\citet{fig06}, \citet{dav07}, and \citet{dav08}, identified two massive clusters in
the red supergiant (RSG) phase, the first containing 14 (see Figure~\ref{RSG1}) and the second containing
26 RSGs, or collectively about 20\% of all such stars known in the Galaxy. The inferred cluster masses are
$\approx$3(10$^4$)~\Msun\ for the former, and $\approx$4(10$^4$)~\Msun\ for the
latter. Interestingly, the two clusters are near each other, located within 1 degree on the sky and 
within 1~kpc along the line of sight at the base of the Scutum-Crux arm. 
The first cluster was first identified as a candidate cluster in the \citet{bic03} 
catalog, and some of the stars in the second cluster had already been identified as RSGs
in \citet{ste90} and \citet{nak01}. Just like in the case of Westerlund~1, we see that
a cluster once thought to be of relatively low mass can turn out to be quite massive on
further inspection. Perhaps there are more massive clusters amongst the many clusters
that have already been identified. 

\begin{figure}
\begin{center}
\end{center}
\caption{GLIMPSE image of RSGC1. W42, to the upper left, is an unrelated star formation region along the line of sight.
See http://www.cis.rit.edu/~dffpci/private/papers/IAU250/Figer2.pdf for high resolution version.\label{RSG1}}
\end{figure}

\subsection{Central cluster}
The Central cluster resides in the central parsec of the Galaxy and 
contains many massive stars formed in the past 10~\Myr\
\citep{bec78,rie78,leb82,for87,all90,kra91,naj94,kra95,naj95,blu95b,gen96,naj97}.
There are at least 80 massive stars in the 
Central cluster \citep{eis05}, including $\approx$50 OB stars on
the main sequence and 30 more evolved massive stars (see Figure~\ref{eckartimage_annot}). 
These young stars appear to be confined to two disks \citep{gen03,tan06}.
There is also a tight collection of a dozen or so B stars (the ``s'' stars) in the central arcsecond,
highlighted in the small box in the figure.  
The formation of so many massive stars in the central parsec
remains as much a mystery now as it was at the time of the first infrared observations of the region.
Most recently, this topic has largely been supplanted by the even more improbable notion that
star formation can occur within a few thousand AU of the supermassive black hole.
See \citet{fig08}, and references therein, for a review of massive star formation and the ``s'' stars
in the Galactic center.

\begin{figure}
\begin{center}
\vskip 0.7truein
\caption{Radio emission from the GC region at 6 cm, adapted by \citet{cot99} from \citet{yus87}. The star
symbols represent the three massive clusters. 
See http://www.cis.rit.edu/~dffpci/private/papers/IAU250/Figer2.pdf for high resolution version.\label{radio}}
\end{center}
\end{figure}

\begin{figure}
\begin{center}
\caption{K-band image of the Central cluster obtained with NAOS/CONICA from \citet{sch06}. The 100 or so brightest
stars in the image are evolved descendants from main sequence O-stars. The central box highlights the ``s'' stars that
are presumably young and massive (\Minit$\approx$20~\Msun). See http://www.cis.rit.edu/~dffpci/private/papers/IAU250/Figer2.pdf for high resolution version.
\label{eckartimage_annot}}
\end{center}
\end{figure}

\subsection{Arches cluster}
The Arches cluster is the densest young cluster in the Galaxy, and it has a
relatively young age \citep{fig02}. Being so young and massive, it contains 
the richest collection of O-stars and WNL stars in any cluster in 
the Galaxy \citep{har94,nag95,fig95t,cot95,cot96,ser98,fig99b,blu01,fig02,fig05}. The WNL stars are
particularly interesting, as they represent the largest collection of the most massive stars (M$>$100~\Msun) in the Galaxy. 
As seen elsewhere, e.g.\ R136 and NGC~3603, these stars are 
still burning hydrogen on the main sequence. 
Given its unique combination of characteristics, 
the cluster is ideal as a testbed for measuring the upper mass cutoff to the IMF
(see Section~\ref{sec:imf}).
The strong stellar winds from the most massive stars in the cluster are detected at radio 
wavelengths \citep{lan01,yus03,lan05,fig02}, and x-ray 
wavelengths \citep{yus02,roc05,wan06}.


\subsection{Quintuplet cluster}
The Quintuplet cluster was originally noted for its five very bright stars \citep{gla90,oku90,nag90}, but
is now known to contain many massive stars \citep{geb94,fig95,tim96,fig99a}.
It is $\approx$4~\Myr\ old and had an initial mass of $>$10$^4$~\Msun\ \citep{fig99a}.
The hot stars in the cluster ionize the nearby ``Sickle'' HII region (see Figure~\ref{radio}).
The Quintuplet is most similar to Westerlund 1 in mass, age, and spectral content.
Some of the stars in the cluster have been detected at x-ray wavelengths 
\citep{law04}, and at radio wavelengths \citep{lan99,lan05}.
Recently, \citet{tut06} convincingly show that the five red stars in the cluster are dusty WC stars,  
characteristic of binary systems containing WCL plus an OB star \citep{tut99}. This may indicate that either the
binary fraction for massive stars is extremely high \citep{nel04}, or only binary massive stars evolve through
the WCL phase \citep{van01}. The Quintuplet cluster also contains two Luminous Blue 
Variables, the Pistol star \citep{har94,fig98,fig99c},
and \FMM362 \citep{fig99a,geb00}. Both stars are extraordinarily luminous (L$>$10$^6$~\Lsun),
yet relatively cool (T$\approx$10$^4$~K), placing them in the ``forbidden zone''
of the Hertzsprung-Russell Diagram, above the Humphreys-Davidson limit \citep{hum94}. They
are also both confirmed photometric and spectroscopic variables \citep{fig99a}.
The
Pistol star is particularly intriguing, in that it is surrounded by one of the most massive (10~\Msun) circumstellar 
ejecta in the Galaxy \citep{fig99c,smi06}. Both stars are spectroscopically 
\citep{fig99a} and photometrically variable \citep{gla01}, as expected for LBVs.

\section{An upper limit to the masses of stars\label{sec:imf}}
Massive star clusters can be useful testing grounds for a variety of
theoretical predictions, e.g.\ the IMF, binary fraction, 
n-body interactions, etc. Although only recently discovered, the currently 
known set of Galactic massive clusters have already yielded an interesting
result regarding the upper limit to the masses of stars. 

Theoretically, one might expect stellar mass to be limited by 
pulsational instabilities \citep{sch59} or radiation pressure \citep{wol87}.
Although stellar evolution models have been computed for massive
stars up masses of 1000~\Msun, no
such stars have ever been observed. Indeed, some of the most famous
``massive stars'' have turned out to be multiple systems. 

Observationally, the problem is difficult because massive stars are rare.
They are formed in small numbers with respect to lower mass stars and 
they only live a few million years. A star formation event must produce 
about 10$^4$~\Msun\ in stars to have a statistically meaningful expectation
of stars with masses greater than 150~\Msun. For example, such a cluster 
should have about three such stars, assuming a Salpeter initial mass
function \citep{sal55} extrapolated to zero probability. Unfortunately, the
cluster must satisfy a number of other criteria in order to use it for
identifying an upper mass cutoff. For instance, it must be young enough so
that its most massive members have not yet exploded as supernovae. Yet, it
must be old enough to be free of natal molecular material. In order to
identify its individual stars, it must be close enough to us. To relate
apparent magnitude to absolute magnitude, it must be at known distance. The
stars must be coeval enough so that the star formation episode that produced
the cluster can be considered to constitute a single event. Finally, the age
must be known relatively accurately.

\begin{figure}
\begin{center}
\includegraphics[height=11cm,angle=90]{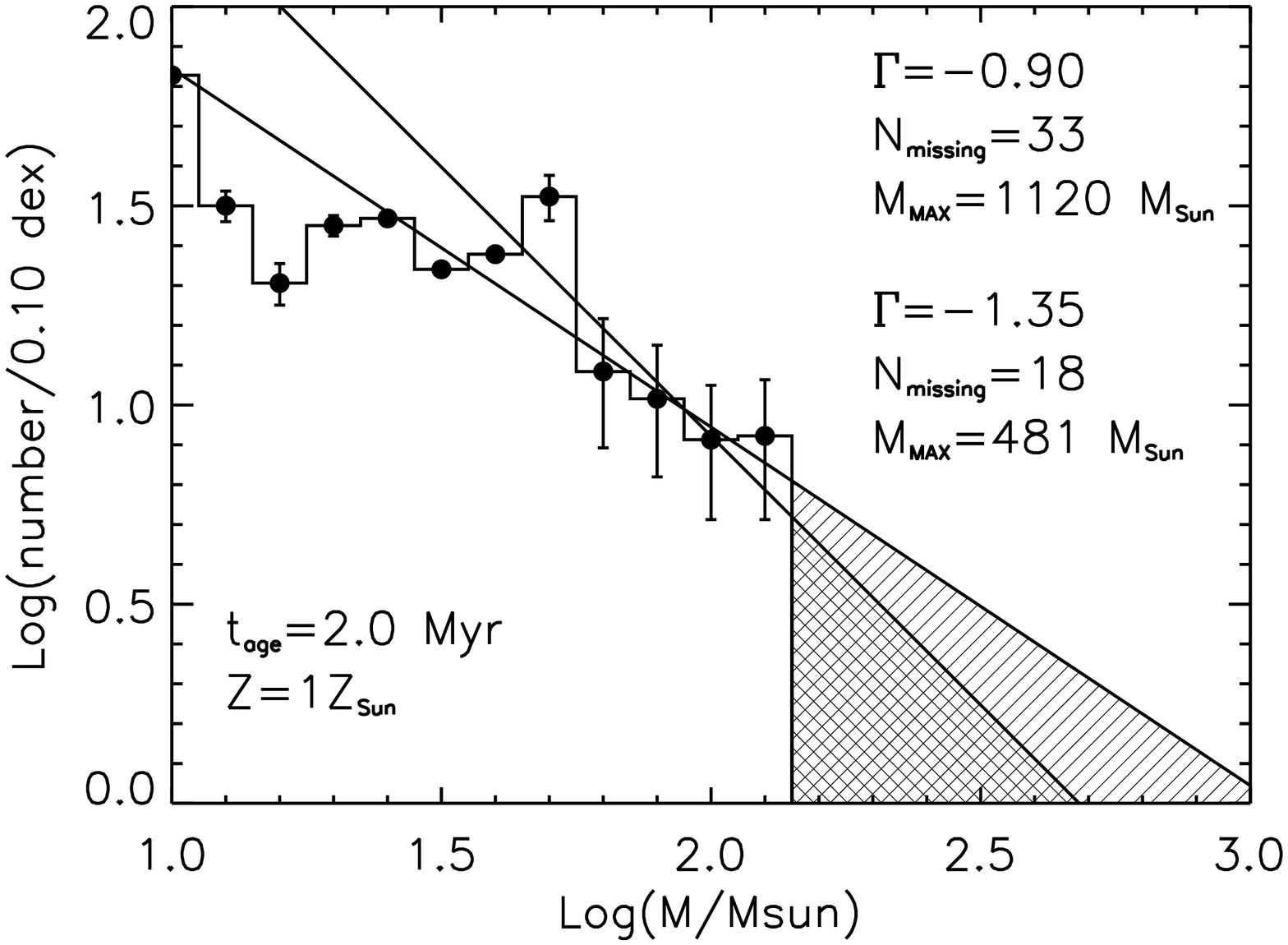}
\end{center}
\caption{Number versus mass for stars in the Arches cluster from \citet{fig05}.
There is a clear deficit of stars
with initial masses greater than 150~\Msun, as seen in the hatched regions, for a 
reasonable range of IMF slopes.\label{upper}}
\end{figure}

It is difficult to satisfy all these criteria, and, indeed, only the Arches
cluster does. Being in the Galactic center happens to be useful in this regard,
as its distance is then relatively well known as compared to other clusters. 
The result for the Arches cluster is shown in Figure~\ref{upper}. In this plot, it
is apparent that there is an absence of stars with masses above about 150~\Msun,
where many are expected, i.e.\ there is an upper mass cutoff. See \citet{wei04},
\citet{wei06} and \citet{oey05} for other arguments for such a cutoff.

\section{The hidden population of massive stellar clusters}
\begin{figure}
\begin{center}
\includegraphics[height=9cm]{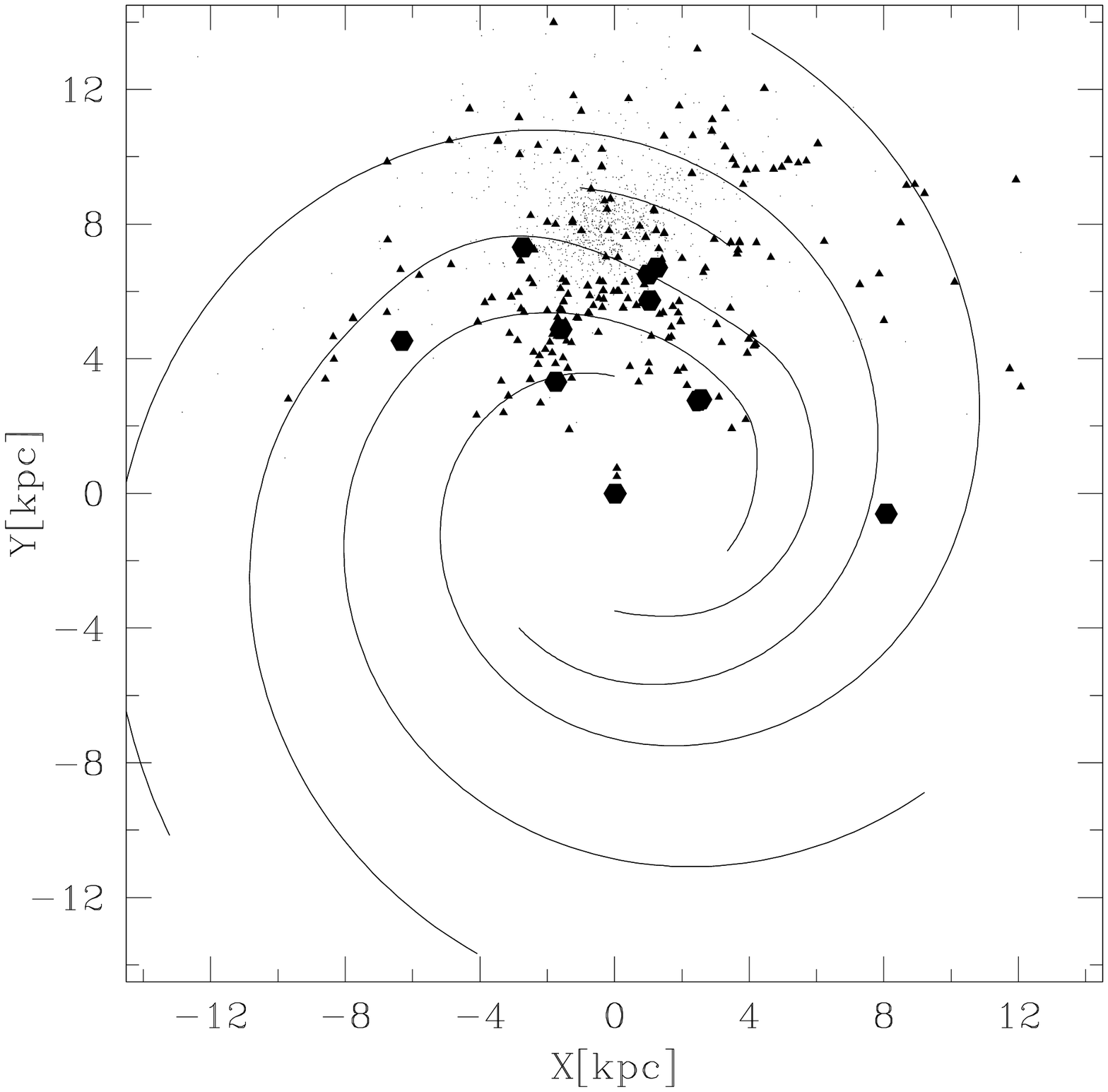}
\end{center}
\caption{Galactic distribution of known young clusters (dots), candidate clusters (triangles), and
verified massive clusters (hexagons). The Galactic center is at (0,0) and the Sun is at (0,8).
Distance for the known young clusters are from \citet{dia02}. Distances for the candidate
clusters were determined by fitting the line-of-sight velocities of nearby HII regions \citep{kuc94} to
the Galactic rotation curve, assuming that these regions are associated with the clusters. This
figure is courtesy of Maria Messineo.\label{messineo}}
\end{figure}

2MASS \citep{skr97} and GLIMPSE \citep{ben03} have heralded a new era in massive star
cluster research. With these surveys, we will be able to probe much further into
the Galactic plane than ever before. We can expect an order of magnitude increase in the
number of known young clusters in the Galaxy as these surveys are further investigated. 
Figure~\ref{messineo} shows the Galactic distribution of known young clusters from WEBDA (dots), 
candidate clusters from \citet{bic03} (triangles), and
the verified massive clusters discussed in this paper, with the addition of several very likely
massive clusters (hexagons). The Galactic center is at (0,0) and the Sun is at (0,8).
The visual clusters are mostly within 3~kpc from the Sun. The infrared clusters are a bit further away;
however, it is clear that the number is highly incomplete 
and that the far side of the Galaxy as well as the central regions must 
contain many new clusters yet to be discovered. We estimate that the total 
number of Galactic stellar clusters should exceed 20,000 \citep{mes08}.

\begin{acknowledgments}
I thank the following individuals for discussions related to this work: 
Ben Davies, Paco Najarro, Rolf Kudritzki, Maria Messineo, Lucy Hadfield, Qingfeng Zhu, and Sungsoo Kim. 
The material in this paper is based upon work supported by NASA
under award No.\ NNG05-GC37G, through the {\it Long Term Space Astrophysics} program.
This research has made use of the
SIMBAD database, Aladin and IDL software packages, and the
GSFC IDL library. This research was performed in the Rochester
Imaging Detector Laboratory with support from a NYSTAR Faculty
Development Program grant. Some of the data presented herein
were obtained at the W. M. Keck Observatory, which is operated
as a scientific partnership among the California Institute of Technology,
the University of California, and the National Aeronautics
and Space Administration. The Observatory was made possible
by the generous financial support of the W. M. Keck Foundation.
\end{acknowledgments}

\begin{discussion}

\discuss{Vanbeveren}{In 1982, I published a study where it was shown (from a theoretical point of view) that the maximum mass of the stars in a cluster depends on the total mass of the cluster and, from the cluster (total) mass function, I proposed a mass function for the maximum mass. This idea has been picked up by Weidner and Kroupa in 2004. Is there any observational evidence that he maximum mass of the stars in a cluster depends on the cluster mass?}

\discuss{Figer}{Some work to answer that question has been done by \citet{oey05}; however, the clusters in that study have very uncertain ages and it is thus not clear if their most massive members have already exploded as supernovae.}

\discuss{Damineli}{Your open cluster candidates do not fit the``radi'' portion of the arms, specifically in the Carina arm. How did you assign distances to the clusters?}

\discuss{Figer}{We considered clusters associated with HII regions for which 
the line-of-velocities are available (Kuchar et al. 1994).
We therefore obtained distances by using the
equations in Burton (1988), assuming  near-distances.}

\discuss{Langer}{You use the argument that no WNE or WC stars are found in the Arches cluster to prove that it is young enough so that the most massive stars have not exploded off yet. Is this challenged by the evidence put forward by Nathan Smith that the most massive stars die as hydrogen-rich LBVs?}

\discuss{Figer}{Nathan Smith's speculation suggests that the most massive stars die as WNh. If this is true, then it is possible that we can no longer see stars 
with initial masses above 150~\Msun\ in the Arches cluster because they have already exploded as supernovae, NOT because they did not form. This would also be 
true for R136 in 30 Dor. One interesting observational test of such a claim is to search for coeval clusters in which there are simultaneously WNh stars 
and more evolved subtypes (presumably from the highest initial masses in the cluster). Perhaps NGC~3603 is promising for such a test where there are WNh stars and a more
evolved subtype, Sher~25; however, one would have to clearly show that these stars are coeval.}

\end{discussion}


\begin{thebibliography}{}


\bibitem[Allen, Hyland, \& Hillier(1990)]{all90} 
\textsc{Allen, D.~A., Hyland, A.~R., \& Hillier, D.~J.} 1990 The source of 
luminosity at the Galactic Centre.  \mnras {\bf 244}, 706 

\bibitem[Barbosa \& Figer(2004)]{bar04} \textsc{Barbosa, C., 
\& Figer, D.} 2004 Top 10 Problems on Massive Stars.  ArXiv Astrophysics 
e-prints, arXiv:astro-ph/0408491 

\bibitem[Becklin et al.(1978)]{bec78} \textsc{Becklin, E.~E., 
Matthews, K., Neugebauer, G., \& Willner, S.~P.} 1978 Infrared observations 
of the galactic center. I - Nature of the compact sources.  \apj {\bf 219}, 
121 

\bibitem[Benjamin et al.(2003)]{ben03} \textsc{Benjamin, 
R.~A., et al.} 2003 GLIMPSE. I. An SIRTF Legacy Project to Map the Inner 
Galaxy.  \pasp {\bf 115}, 953 



\bibitem[Bica et al.(2003)]{bic03} \textsc{Bica, E., Dutra, 
C.~M., Soares, J., \& Barbuy, B.} 2003 New infrared star clusters in the 
Northern and Equatorial Milky Way with 2MASS.  \aap {\bf 404}, 223 

\bibitem[Blum et al.(2001)]{blu01} \textsc{Blum, R.~D., 
Schaerer, D., Pasquali, A., Heydari-Malayeri, M., Conti, P.~S., \& Schmutz, 
W.} 2001 2 Micron Narrowband Adaptive Optics Imaging in the Arches Cluster.  
\aj {\bf 122}, 1875 

\bibitem[Blum, Sellgren, \& Depoy(1995b)]{blu95b} 
\textsc{Blum, R.~D., Sellgren, K., \& Depoy, D.~L.} 1995b Discovery of a 
possible Wolf-Rayet star at the galactic center.  \apjl {\bf 440}, L17 

\bibitem[Blum, Sellgren, \& Depoy(1996a)]{blu96a} \textsc{Blum, 
R.~D., Sellgren, K., \& Depoy, D.~L.} 1996a Really Cool Stars at the 
Galactic Center.  \aj {\bf 112}, 1988 

\bibitem[Brandner et al.(2008)]{bra08} \textsc{Brandner, W., 
Clark, J.~S., Stolte, A., Waters, R., Negueruela, I., \& Goodwin, S.~P.} 
2008 Intermediate to low-mass stellar content of Westerlund 1.  \aap {\bf 
478}, 137 



\bibitem[Clark et al.(2005)]{cla05} \textsc{Clark, J.~S., 
Negueruela, I., Crowther, P.~A., \& Goodwin, S.~P.} 2005 On the massive 
stellar population of the super star cluster Westerlund 1.  \aap {\bf 434}, 
949 

\bibitem[Cotera(1995)]{cot95} \textsc{Cotera, A.~S.} 1995 
Stellar Ionization of the Thermal Emission Regions in the Galactic Center.  
Ph.D.~Thesis

\bibitem[Cotera et al.(1996)]{cot96} \textsc{Cotera, A.~S., 
Erickson, E.~F., Colgan, S.~W.~J., Simpson, J.~P., Allen, D.~A., \& Burton, 
M.~G.} 1996 The Discovery of Hot Stars near the Galactic Center Thermal 
Radio Filaments.  \apj {\bf 461}, 750 

\bibitem[Cotera et al.(1999)]{cot99} \textsc{Cotera, A.~S., 
Simpson, J.~P., Erickson, E.~F., Colgan, S.~W.~J., Burton, M.~G., \& Allen, 
D.~A.} 1999 Isolated Hot Stars in the Galactic Center Vicinity.  \apj {\bf 
510}, 747 

\bibitem[Crowther et al.(2006)]{cro06} \textsc{Crowther, P.A., Hadfield, L. J., Clark, J. S., Negueruela, I., \& Vacca, W. D.} 
2006 A census of the Wolf-Rayet content in Westerlund 1 from near-infrared imaging and spectroscopy. 
ArXiv Astrophysics e-prints, 
arXiv:astro-ph/0608356 

\bibitem[Davies et al.(2007)]{dav07} \textsc{Davies, B., 
Figer, D.~F., Kudritzki, R.-P., MacKenty, J., Najarro, F., \& Herrero, A.} 
2007 A Massive Cluster of Red Supergiants at the Base of the Scutum-Crux 
Arm.  \apj {\bf 671}, 781 

\bibitem[Davies et al.(2008)]{dav08} \textsc{Davies, B., 
Figer, D.~F., Law, C.~J., Kudritzki, R.-P., Najarro, F., Herrero, A., \& 
MacKenty, J.~W.} 2007 The cool supergiant population of the massive young 
star cluster RSGC1.  ArXiv e-prints {\bf 711}, arXiv:0711.4757 

\bibitem[Dias et al.(2002)]{dia02} \textsc{Dias, W.~S., 
Alessi, B.~S., Moitinho, A., \& L{\'e}pine, J.~R.~D.} 2002 New catalogue of 
optically visible open clusters and candidates.  \aap {\bf 389}, 871 




\bibitem[Eisenhauer et al.(2005)]{eis05} 
\textsc{Eisenhauer, F., et al.} 2005 SINFONI in the Galactic Center: Young 
Stars and Infrared Flares in the Central Light-Month.  \apj {\bf 628}, 246 

\bibitem[Figer(1995)]{fig95t} \textsc{Figer, D.~F.} 1995 A 
Search for Emission-Line Stars Near the Galactic Center.  Ph.D.~Thesis.



\bibitem[Figer(2005)]{fig05} \textsc{Figer, D.~F.} 2005 An 
upper limit to the masses of stars.  \nat {\bf 434}, 192 

\bibitem[Figer(2008)]{fig08} \textsc{Figer, D.~F.} 2008 Massive Star Formation in the Galactic Center.  
Massive Stars: From Pop III and GRBs to the Milky Way, in press


\bibitem[Figer et al.(2002)]{fig02} 
	\textsc{Figer, D.~F.~et al.} 
	2002 Massive Stars in the Arches Cluster. \apj \textbf{581}, 258 




\bibitem[Figer et al.(1999b)]{fig99b} 
	\textsc{Figer, D.\ F., Kim, S.\ S., Morris, M., Serabyn, E., Rich, R.\ M., \& McLean, I.\ S.} 
	1999b
	{HST/NICMOS Observations of Massive Stellar Clusters Near the Galactic Center.}
	\apj \textbf{525}, 750.

\bibitem[Figer et al.(2006)]{fig06} \textsc{Figer, D.~F., 
MacKenty, J.~W., Robberto, M., Smith, K., Najarro, F., Kudritzki, R.~P., \& 
Herrero, A.} 2006 Discovery of an Extraordinarily Massive Cluster of Red 
Supergiants.  \apj {\bf 643}, 1166 

\bibitem[Figer, McLean, \& Morris(1995)]{fig95} 
\textsc{Figer, D.~F., McLean, I.~S., \& Morris, M.} 1995 Two New Wolf-Rayet 
Stars and a Luminous Blue Variable Star in the Quintuplet (AFGL 2004) near 
the Galactic Center.  \apjl {\bf 447}, L29 

\bibitem[Figer et al.(1999a)]{fig99a} 
	\textsc{Figer, D. F., McLean, I. S., \& Morris, M.} 
	1999a, Massive Stars in the Quintuplet Cluster. \apj \textbf{514}, 202 

\bibitem[Figer et al.(1999c)]{fig99c} \textsc{Figer, D.~F., 
Morris, M., Geballe, T.~R., Rich, R.~M., Serabyn, E., McLean, I.~S., 
Puetter, R.~C., \& Yahil, A.} 1999c High-Resolution Infrared Imaging and 
Spectroscopy of the Pistol Nebula: Evidence for Ejection.  \apj {\bf 525}, 
759 


\bibitem[Figer et al.(2005)]{fig05a} \textsc{Figer, D.~F., 
Najarro, F., Geballe, T.~R., Blum, R.~D., \& Kudritzki, R.~P.} 2005 Massive 
Stars in the SGR 1806-20 Cluster.  \apjl {\bf 622}, L49 


\bibitem[Figer et al.(1998)]{fig98} \textsc{Figer, D.~F., 
Najarro, F., Morris, M., McLean, I.~S., Geballe, T.~R., Ghez, A.~M., \& 
Langer, N.} 1998 The Pistol star.  \apj {\bf 506}, 384 


\bibitem[Forrest et al.(1987)]{for87} \textsc{Forrest, 
W.~J., Shure, M.~A., Pipher, J.~L., \& Woodward, C.~E.} 1987 Brackett Alpha 
Images.  AIP Conf.~Proc.~155: The Galactic Center {\bf 155}, 153 



\bibitem[Geballe et al.(1994)]{geb94} \textsc{Geballe, 
T.~R., Genzel, R., Krabbe, A., Krenz, T., \& Lutz, D.} 1994 Spectra of a 
Remarkable Class of Hot Stars in the Galactic Center.  ASSL Vol.~190: 
Astronomy with Arrays, The Next Generation, 73 

\bibitem[Geballe, Najarro, \& Figer(2000)]{geb00} 
\textsc{Geballe, T.~R., Najarro, F., \& Figer, D.~F.} 2000 A Second 
Luminous Blue Variable in the Quintuplet Cluster.  \apjl {\bf 530}, L97 

\bibitem[Genzel et al.(2003)]{gen03} \textsc{Genzel, R., et 
al.} 2003 The Stellar Cusp around the Supermassive Black Hole in the 
Galactic Center.  \apj {\bf 594}, 812 



\bibitem[Genzel et al.(1996)]{gen96} \textsc{Genzel, R., 
Thatte, N., Krabbe, A., Kroker, H., \& Tacconi-Garman, L.~E.} 1996 The Dark 
Mass Concentration in the Central Parsec of the Milky Way.  \apj {\bf 472}, 
153 



\bibitem[Glass et al.(2001)]{gla01} \textsc{Glass, I.~S., 
Matsumoto, S., Carter, B.~S., \& Sekiguchi, K.} 2001 Large-amplitude 
variables near the Galactic Centre.  \mnras {\bf 321}, 77 

\bibitem[Glass, Moneti, \& Moorwood(1990)]{gla90} 
\textsc{Glass, I.~S., Moneti, A., \& Moorwood, A.~F.~M.} 1990 Infrared 
images and photometry of the cluster near G 0.15 - 0.05.  \mnras {\bf 242}, 
55P 

\bibitem[Groh et al.(2006)]{gro06} \textsc{Groh, J.~H., 
Damineli, A., Teodoro, M., \& Barbosa, C.~L.} 2006 Detection of additional 
Wolf-Rayet stars in the starburst cluster Westerlund 1 with SOAR.  ArXiv 
Astrophysics e-prints, arXiv:astro-ph/0606498 

\bibitem[Harayama(2007)]{har07} \textsc{Harayama, Y.} 2007 The IMF of the massive star-forming region NGC 3603 from NIR adaptive optics observations
. PhD Thesis, LMU München


\bibitem[Harris et al.(1994)]{har94} \textsc{Harris, A.~I., 
Krenz, T., Genzel, R., Krabbe, A., Lutz, D., Politsch, A., Townes, C.~H., 
\& Geballe, T.~R.} 1994 Spectroscopy of the Galactic Center Arches Region: 
Evidence for Massive Star Formation.  NATO ASIC Proc.~445: The Nuclei of 
Normal Galaxies: Lessons from the Galactic Center, 223 





\bibitem[Kim et al.(2006)]{kim06} \textsc{Kim, S.~S., Figer, 
D.~F., Kudritzki, R.~P., \& Najarro, F.} 2006 The Arches Cluster Mass 
Function.  \apjl {\bf 653}, L113 



\bibitem[Krabbe et al.(1995)]{kra95} \textsc{Krabbe, A., et 
al.} 1995 The Nuclear Cluster of the Milky Way: Star Formation and Velocity 
Dispersion in the Central 0.5 Parsec.  \apjl {\bf 447}, L95 

\bibitem[Krabbe et al.(1991)]{kra91} \textsc{Krabbe, A., 
Genzel, R., Drapatz, S., \& Rotaciuc, V.} 1991 A cluster of He I 
emission-line stars in the Galactic center.  \apjl {\bf 382}, L19 

\bibitem[Kuchar \& Bania(1994)]{kuc94} \textsc{Kuchar, T.~A., 
\& Bania, T.~M.} 1994 Kinematic distances of Galactic H II regions from H I 
absorption studies.  \apj {\bf 436}, 117 

\bibitem[Lang et al.(1999)]{lan99} 
\textsc{Lang, C.~C., Figer, D.~F., Goss, W.~M., \& Morris, M.} 1999 Radio 
Detections of Stellar Winds from the Pistol star and Other Stars in the 
Galactic Center Quintuplet Cluster.  \aj {\bf 118}, 2327 

\bibitem[Lang, Goss, \& Rodr{\'{\i}}guez(2001)]{lan01} 
\textsc{Lang, C.~C., Goss, W.~M., \& Rodr{\'{\i}}guez, L.~F.} 2001 Very 
Large Array Detection of the Ionized Stellar Winds Arising from Massive 
Stars in the Galactic Center Arches Cluster.  \apjl {\bf 551}, L143 


\bibitem[Lang et al.(2005)]{lan05} \textsc{Lang, C.~C., 
Johnson, K.~E., Goss, W.~M., \& Rodr{\'{\i}}guez, L.~F.} 2005 Stellar Winds 
and Embedded Star Formation in the Galactic Center Quintuplet and Arches 
Clusters: Multifrequency Radio Observations.  \aj {\bf 130}, 2185 

\bibitem[Law \& Yusef-Zadeh(2004)]{law04} \textsc{Law, C., 
\& Yusef-Zadeh, F.} 2004 X-Ray Observations of Stellar Clusters Near the 
Galactic Center.  \apj {\bf 611}, 858 

\bibitem[Lebofsky, Rieke, \& Tokunaga(1982)]{leb82} 
\textsc{Lebofsky, M.~J., Rieke, G.~H., \& Tokunaga, A.~T.} 1982 M 
supergiants and star formation at the galactic center.  \apj {\bf 263}, 736 


\bibitem[Messineo \& Figer(2008)]{mes08} 
	\textsc{Messineo, M. \& Figer, D.~F.} 
	2003, Massive Stars as Cosmic Engines. F. Bresolin, P.A. Crowther \& J. Puls, eds., IAU Symposium, \textbf{250}, these proceedings








\bibitem[Muno et al.(2006)]{mun06} \textsc{Muno, M.~P., et 
al.} 2006 A Neutron Star with a Massive Progenitor in Westerlund 1.  \apjl 
{\bf 636}, L41 



\bibitem[Nagata et al.(1995)]{nag95} \textsc{Nagata, T., 
Woodward, C.~E., Shure, M., \& Kobayashi, N.} 1995 Object 17: Another 
cluster of emission-line stars near the galactic center.  \aj {\bf 109}, 
1676 

\bibitem[Nagata et al.(1990)]{nag90} \textsc{Nagata, T., 
Woodward, C.~E., Shure, M., Pipher, J.~L., \& Okuda, H.} 1990 AFGL 2004 - 
an infrared quintuplet near the Galactic center.  \apj {\bf 351}, 83 

\bibitem[Najarro(1995)]{naj95} \textsc{Najarro, F.} 1995 
Quantitative Optical and Infrared Spectroscopy of Extreme Luminous Blue 
Supergiants.  Ph.D.~Thesis 


\bibitem[Najarro et al.(1994)]{naj94} \textsc{Najarro, F., 
Hillier, D.~J., Kudritzki, R.~P., Krabbe, A., Genzel, R., Lutz, D., 
Drapatz, S., \& Geballe, T.~R.} 1994 The nature of the brightest galactic 
center HeI emission line star.  \aap {\bf 285}, 573 

\bibitem[Najarro et al.(1997)]{naj97} \textsc{Najarro, F., 
Krabbe, A., Genzel, R., Lutz, D., Kudritzki, R.~P., \& Hillier, D.~J.} 1997 
Quantitative spectroscopy of the HeI cluster in the Galactic center..  \aap 
{\bf 325}, 700 

\bibitem[Nakaya et al.(2001)]{nak01} \textsc{Nakaya, H., 
Watanabe, M., Ando, M., Nagata, T., \& Sato, S.} 2001 A Highly Reddened 
Star Cluster Embedded in the Galactic Plane.  \aj {\bf 122}, 876 

\bibitem[Negueruela \& Clark(2005)]{neg05} 
\textsc{Negueruela, I., \& Clark, J.~S.} 2005 Further Wolf-Rayet stars in 
the starburst cluster Westerlund 1.  \aap {\bf 436}, 541 

\bibitem[Nelan et al.(2004)]{nel04} \textsc{Nelan, E.~P., 
Walborn, N.~R., Wallace, D.~J., Moffat, A.~F.~J., Makidon, R.~B., Gies, 
D.~R., \& Panagia, N.} 2004 Resolving OB Systems in the Carina Nebula with 
the Hubble Space Telescope Fine Guidance Sensor.  \aj {\bf 128}, 323 

\bibitem[Oey \& Clarke(2005)]{oey05} \textsc{Oey, M.~S., \& 
Clarke, C.~J.} 2005 Statistical Confirmation of a Stellar Upper Mass Limit.  
\apjl {\bf 620}, L43 

\bibitem[Okuda et al.(1990)]{oku90} \textsc{Okuda, H., et al.} 1990 An infrared quintuplet near the Galactic center.  \apj {\bf 351}, 89 





\bibitem[Rieke, Telesco, \& Harper(1978)]{rie78} 
\textsc{Rieke, G.~H., Telesco, C.~M., \& Harper, D.~A.} 1978 The infrared 
emission of the Galactic center.  \apj {\bf 220}, 556 

\bibitem[Rockefeller et al.(2005)]{roc05} 
\textsc{Rockefeller, G., Fryer, C.~L., Melia, F., \& Wang, Q.~D.} 2005 
Diffuse X-Rays from the Arches and Quintuplet Clusters.  \apj {\bf 623}, 
171 

\bibitem[Salpeter(1955)]{sal55} 
	\textsc{Salpeter, E. E.} 
	1955 The Luminosity Function and Stellar Evolution. \apj \textbf{121}, 161


\bibitem[Schwarzschild \& {\ H\"a}rm(1959)]{sch59} 
\textsc{Schwarzschild, M., {\ H\"a}rm, R.} 1959 On the Maximum Mass of 
Stable Stars.  \apj {\bf 129}, 637 



\bibitem[Sch{\"o}del et al.(2006)]{sch06} 
\textsc{Sch{\"o}del, R., et al.} 2006 \aap, in preparation


\bibitem[Serabyn, Shupe, \& Figer(1998)]{ser98} 
\textsc{Serabyn, E., Shupe, D., \& Figer, D.~F.} 1998 An extraordinary 
cluster of massive stars near the centre of the Milky Way..  \nat {\bf 
394}, 448 

\bibitem[Skinner et al.(2006)]{ski06} \textsc{Skinner, 
S.~L., Simmons, A.~E., Zhekov, S.~A., Teodoro, M., Damineli, A., \& Palla, 
F.} 2006 A Rich Population of X-Ray-emitting Wolf-Rayet Stars in the 
Galactic Starburst Cluster Westerlund 1.  \apjl {\bf 639}, L35 

\bibitem[Skrutskie et al.(1997)]{skr97} \textsc{Skrutskie, 
M.~F., et al.} 1997 The Two Micron All Sky Survey (2MASS): Overview and 
Status.  The Impact of Large Scale Near-IR Sky Surveys {\bf 210}, 25 



\bibitem[Smith(2006)]{smi06} \textsc{Smith, N.} 2006 Eruptive 
Mass Loss in Very Massive Stars and Population III Stars.  ArXiv 
Astrophysics e-prints, arXiv:astro-ph/0607457 

\bibitem[Stephenson(1990)]{ste90} \textsc{Stephenson, C.~B.} 
1990 A possible new and very remote galactic cluster.  \aj {\bf 99}, 1867 






\bibitem[Tanner et al.(2006)]{tan06} \textsc{Tanner, A., et al.} 
2006 High Spectral Resolution Observations of the Massive Stars in the 
Galactic Center.  \apj {\bf 641}, 891 

\bibitem[Timmermann et al.(1996)]{tim96} 
\textsc{Timmermann, R., Genzel, R., Poglitsch, A., Lutz, D., Madden, S.~C., 
Nikola, T., Geis, N., \& Townes, C.~H.} 1996 Far-Infrared Observations of 
the Radio Arc (Thermal Arches) in the Galactic Center.  \apj {\bf 466}, 242 


\bibitem[Tuthill, Monnier, \& Danchi(1999)]{tut99} 
\textsc{Tuthill, P.~G., Monnier, J.~D., \& Danchi, W.~C.} 1999 A dusty 
pinwheel nebula around the massive star WR 104..  \nat {\bf 398}, 487 



\bibitem[Tuthill et al.(2006)]{tut06}
	\textsc{Tuthill, P., Monnier, J., Tanner, A., Figer, D., \& Ghez, A.} 
	2006
	{``Pinwheels'' discovered in the Quintuplet cluster.}
	\emph{Science} {\bf 313}, 935.


\bibitem[van der Hucht(2001)]{van01} \textsc{van der Hucht, 
K.~A.} 2001 The VIIth catalogue of galactic Wolf-Rayet stars.  New 
Astronomy Review {\bf 45}, 135 


\bibitem[Wang, Dong, \& Lang(2006)]{wan06} \textsc{Wang, 
Q.~D., Dong, H., \& Lang, C.} 2006 The Interplay between Star Formation and 
the Nuclear Environment of our Galaxy: Deep X-ray Observations of the 
Galactic Center Arches and Quintuplet Clusters.  ArXiv Astrophysics 
e-prints, arXiv:astro-ph/0606282 

\bibitem[Weidner \& Kroupa(2004)]{wei04} \textsc{Weidner, C., 
\& Kroupa, P.} 2004 Evidence for a fundamental stellar upper mass limit 
from clustered star formation.  \mnras {\bf 348}, 187 


\bibitem[Weidner \& Kroupa(2006)]{wei06} \textsc{Weidner, C., 
\& Kroupa, P.} 2006 The maximum stellar mass, star-cluster formation and 
composite stellar populations.  \mnras {\bf 365}, 1333 




\bibitem[Wolfire \& Cassinelli(1987)]{wol87} \textsc{Wolfire, 
M.~G., \& Cassinelli, J.~P.} 1987 Conditions for the formation of massive 
stars.  \apj {\bf 319}, 850 


\bibitem[Yusef-Zadeh et al.(2002)]{yus02} 
\textsc{Yusef-Zadeh, F., Law, C., Wardle, M., Wang, Q.~D., Fruscione, A., 
Lang, C.~C., \& Cotera, A.} 2002 Detection of X-Ray Emission from the 
Arches Cluster near the Galactic Center.  \apj {\bf 570}, 665 

\bibitem[Yusef-Zadeh \& Morris(1987)]{yus87} 
\textsc{Yusef-Zadeh, F., \& Morris, M.} 1987 Structural details of the 
Sagittarius A complex - Evidence for a large-scale poloidal magnetic field 
in the Galactic center region.  \apj {\bf 320}, 545 



\bibitem[Yusef-Zadeh et al.(2003)]{yus03} 
\textsc{Yusef-Zadeh, F., Nord, M., Wardle, M., Law, C., Lang, C., \& Lazio, 
T.~J.~W.} 2003 Nonthermal Emission from the Arches Cluster (G0.121+0.017) 
and the Origin of gamma-ray Emission from 3EG J1746-2851.  \apjl {\bf 590}, 
L103 




\end{thebibliography}
\end{document}